\documentclass{article}
\usepackage{subcaption}
\usepackage{graphicx} % Required for inserting images
\usepackage{multicol}
\usepackage{tcolorbox}
\usepackage{xcolor}
\usepackage{hyperref}
\hypersetup{colorlinks,allcolors=black}
\definecolor{mypink1}{rgb}{0.9, 0.5, 0.5}
\usepackage[backend=biber,style=numeric,sorting=none]{biblatex} 
\usepackage{amsmath}
\usepackage{tikz}
\usetikzlibrary{shapes.geometric, arrows}
\usetikzlibrary{calc}
\tikzstyle{startstop} = [rectangle, rounded corners, minimum width=3cm, minimum height=1cm,text centered, draw=black, fill=red!30]
\tikzstyle{io} = [trapezium, trapezium left angle=70, trapezium right angle=110, minimum width=3cm, minimum height=1cm, text centered, draw=black, fill=blue!30]
\tikzstyle{process} = [rectangle, minimum width=3cm, minimum height=1cm, text centered, draw=black, fill=orange!30]
\tikzstyle{decision} = [diamond, minimum width=3cm, minimum height=1cm, text centered, draw=black, fill=green!30]
\tikzstyle{arrow} = [thick,->,>=stealth]
\tikzstyle{process} = [rectangle, minimum width=3cm, minimum height=1cm, text centered, text width=3cm, draw=black, fill=orange!30]
\tikzstyle{io} = [trapezium, 
trapezium stretches=true, 
trapezium left angle=70, 
trapezium right angle=110, 
minimum width=3cm, 
minimum height=1cm, text centered, 
draw=black, fill=blue!30]
\usepackage{geometry} 
\title{
        \centering
        Comparing Methods of Expert Elicitation for Treatment Effect or Borrowing Parameters in Standard and Rare Disease Clinical Trials: A Systematic Mapping Study
}
\author{Laura M. Morgan$^{1,3}$, James M. S. Wason$^1$, Kevin J. Wilson$^2$, Nina Wilson$^1$}
\date{\small{March 2025}}
\begin{document}
\maketitle
\noindent \emph{\small{$^1$Biostatistics Research Group, Newcastle University. 
$^2$School of Mathematics, Statistics and Physics, Newcastle University. 
$^3$l.m.morgan2@newcastle.ac.uk}}

\section*{\centering Abstract}
\textbf{\emph{Background:}} Expert elicitation is an invaluable tool for gaining insights into the degree of clinical knowledge surrounding parameters of interest when designing, or supplementing trial data when analysing, a clinical trial. Elicitation is considered particularly useful in cases where limited data are available, such as in rare diseases.  
\\

\noindent\textbf{\emph{Methods and objective:}} This study aims to identify methods of expert elicitation and aggregation for treatment effect or difference parameters that are used in the design or analysis stages of clinical trials. A comprehensive review of statistical and non-statistical literature was conducted by database searching, and reference list screening of older, relevant literature reviews. The search took place in October 2024 and identified 366 potentially relevant publications. Of these, 126 were selected for full-text review based on review of titles and abstracts, and 41 publications were deemed eligible for inclusion after a full reading. For each included publication, data were extracted on methods of elicitation and aggregation, the types of parameters elicited, the resulting distributions (parametric versus nonparametric), the number of experts used, and any training provided to experts. Publication  characteristics such as contribution type, journal type, and application to the rare disease setting was also noted.
\\

\noindent \textbf{\emph{Results:}} A narrative description of the selected publications was produced, detailing 6 unique methods for expert elicitation and 10 unique methods for aggregation. We discuss the most popular methods used across standard and rare disease clinical trials, along with any strengths and limitations.
\\

\noindent \textbf{\emph{Conclusion:}} Overall, there is no formal framework for expert elicitation in clinical trials, and more general methods are applied with little consideration into the specific context and trial designs. For instance, despite the unique practical challenges associated with rare disease trials, there have been no specific adaptations made to elicitation methodologies to better suit this context. This review identifies the methodological gaps in current practice, providing a foundation for future development.

\section{Introduction}
Historically, frequentist methods have dominated clinical trials. However, computational advances have facilitated the growth of Bayesian methods, allowing for the incorporation of evidence additional to data collected in the trial. Bayesian inference seeks to assign prior distributions to parameters based on external evidence, quantifying initial beliefs and uncertainty about parameter values\textsuperscript{\cite{stefan2022expert}}. These distributions can be used to document clinical knowledge or inform a sample size calculation, or combined with trial data through Bayes' Theorem for interim study monitoring or final analyses\textsuperscript{\cite{johnson2010methods}}.
\\

\noindent %Choosing an informative prior distribution, which incorporate relevant historical data or expert opinion, can be difficult. 
Prior distributions that incorporate relevant external evidence are considered `informative', but choosing them comes with considerable difficulty. One common approach is to use prior elicitation, which involves gathering information from a group of experts in the field and guiding them through the process of expressing their judgements in the form of probability distributions\textsuperscript{\cite{stefan2022expert,mikkola2024prior}}. A protocol that is commonly employed is the SHeffield ELicitation Framework (SHELF)\textsuperscript{\cite{o2013shelf}}, which includes documents, templates and software for preparing for and conducting elicitation. Nonetheless, despite dating back to the 1960s, there is still a lack of practical tools and standard protocols that can routinely be used as part of the process of elicitation in trials\textsuperscript{\cite{mikkola2024prior}}, particularly when considering rare diseases. 
\\

\noindent In the rare disease context, aggregating expert judgement into a single prior distribution for a Bayesian analysis can be particularly challenging. The rarity of these diseases often means that even experts could have limited experience, potentially leading to greater variation both within, and between expert judgements and, hence, potentially more extreme judgements. The potential for greater uncertainty in the experts' judgements could lead to more diffuse priors from individuals, leading to a relatively non-informative aggregated distribution. Since trials in common diseases are sufficiently powered to detect treatment effects, the data is able to dominate a non-informative prior to still result in meaningful conclusions. However, in rare disease trials, the small sample size causes issues with powering the trial, and so the external information gathered in prior elicitation can be crucial to determining the conclusions of the trial. Furthermore, since rare diseases are often severe and life-threatening, experts may genuinely struggle to quantify key outcomes such as life expectancy or timescales for remission or mortality. This issue becomes especially pronounced in treatments aimed at managing rather than curing the disease. For example, it is less intuitive to quantify whether a treatment aimed at slowing neural decline is effective, compared to whether an intervention results in the complete recovery of a patient. In addition, when untreated disease progression is rapid, it can be particularly challenging to estimate outcomes for control groups, such as expected survival without intervention.
\\

\noindent In this study, we conduct a systematic review of publications that propose or use methods of expert elicitation to obtain distributions on treatment effect or borrowing parameters in the design or analysis of clinical trials. Within `treatment effect parameters', we consider approaches that elicit experts’ expectations about the effectiveness of the treatment(s) in the trial, while borrowing parameters relate to the degree of commensurability between different baskets or between historical and current trial data, enabling the borrowing of data and information across sources in the analysis.
\\

\noindent We summarise the current methodology, its strengths, and its limitations, with focuses on aggregation methods, and on trials in rare diseases that elicit expert judgement on how to use data from related studies. %, and how expert knowledge is elicited in these cases.
Consideration is given to how such trials report their elicitation strategies to identify any inconsistencies in application. Section 2 outlines the methods and search strategy used to identify relevant publications. In Section 3, we present the results, including summaries of publication characteristics, training provided to experts, methods of elicitation, which variables are elicited and how, and methods of aggregation, making comparisons across rare and non-rare disease trials. Section 4 provides a summary of our findings, discusses methodological limitations, and offers recommendations for future research in expert elicitation. 

\section{Methods}
\subsection{Search strategy}
Eligible publications were identified through database searching of PubMed, Web of Science and Scopus, as well as through screening of the reference lists of related reviews. The search took place in October 2024, with no restrictions on the age of the study. The studies were all published in English. 
\\

\noindent Search terms were informed by the authors' knowledge of the area, and adapted based on quantity and type of search results. The search string was set to be general to ensure all relevant results could be found: ``((Bayesian) AND (clinical trial*)) AND (elicitation)". As an exercise in identifying specific studies in rare diseases, the search term was extended to ``(((Bayesian) AND (clinical trial*)) AND (elicitation)) AND (rare disease*)", but this was deemed redundant as all such studies appeared in the initial search. 
\\

\noindent A simple \tt R \normalfont function was designed to eliminate duplicate documents based on their unique Digital Object Identifiers (DOIs), before the titles and abstracts of the resulting list were screened and ineligible publications removed. The remaining publications were subject to a full-text review against inclusion-exclusion criteria by one author (LMM). Of those deemed eligible for full-text review, 10\% were then independently checked by reviewers (JMSW, KJW, and NW) to ensure agreement across all authors. The inclusion-exclusion criteria are defined in Section \ref{criteria}.

\subsection{Inclusion and exclusion criteria}\label{criteria}
To constrain the scope of this review, publications needed to either address or apply elicitation of a prior probability distribution from experts in the field of study, i.e. relevant disease medical professionals, and academics only. That is, any publications defining their prior using exclusively historical data were excluded, unless a prior distribution on the borrowing parameter defining commensurability between the historical and new trial data was elicited from experts. 
\\

\noindent Constraints were placed on the type of parameters elicited - we were only interested in publications that elicited prior distributions on treatment parameters (for example, the efficacy of the intervention), or parameters that enable borrowing (essentially asking ``how similar are different subsets of data?", or ``how relevant are historical data to this trial?"). This led to the exclusion of publications eliciting upper and lower limits of acceptable toxicity in dose-finding studies, as well as decision-theoretic publications that consider `Go/No Go' decisions or eliciting utility weights. Included publications also needed to be concerned with the design or analysis stages of human clinical trials, and describe more than one stage of the elicitation process. This means any publications pertaining to Health Technology Assessments (HTAs), cost-effectiveness analyses, or long-term survival analysis and extrapolation were excluded. 
\\

\noindent In summary, this means that only publications describing elicitation of expert judgement of treatment or borrowing parameters, including uncertainty, to obtain a distribution that could be used in the design or analysis stages of a human clinical trial in sufficient detail to facilitate reasonable comprehension and possible reproduction of the method were deemed pertinent to this study. 
\\

\noindent Publications making it through screening were then further classified by whether they relate to rare diseases, or not. %Citations making it through all previous screening were then subject to a `dummy' inclusion criteria pertaining to any mention of rare diseases. 
This aimed to identify publications eligible for in-depth data extraction for the discussion of methods used in this setting, and a tabular summary of these publications can be found within Additional file 1. 

\subsection{Data extraction and methodologic assessment}
For any included publication, the following data were extracted: whether the approach was parametric or nonparametric, the resulting family of parameteric prior distribution(s), the elicited parameters, the number of experts involved, any training provided to experts, the method of elicitation, the method of aggregation, and the software used. Characteristics of the publications were noted, including their contribution type (methodological/theoretical versus applied), and Journal Citation Report Classification to understand the scope of elicitation methodology. If applicable, the trial phase was reported. 
\\

\noindent For rare disease publications, explicit comparisons were made to consider how methods of elicitation and aggregation reported in the full-text screening had been applied to these trials. While some attention was given to the selection of experts, focus was primarily on how they were subsequently trained and prepared for the elicitation process, given the tendency and requirement for informative priors to dominate the analysis when sample size is low. Due to the potential for greater variation both within individual expert prior distributions, and between the experts, we considered any differences in the aggregation methods between the two disease settings.

\section{Results}
\subsection{Search results and exclusions}
\begin{figure}
\begin{tikzpicture}[node distance = 2cm]
    \node[startstop, draw, text width=7cm, align=center] (total) 
    {\textbf{$\boldsymbol{N=366}$ publications identified:} \\[0.5em]
    PubMed: $n_1=157$ \\ 
    Web of Science: $n_2=106$ \\ 
    Scopus: $n_3=93$ \\
    Screening of reference lists: $n_4=10$ };

    \node[startstop, draw, text width=5cm, align=center, below of=total,yshift=-5cm] (full) 
   {$\boldsymbol{M=126}$ \textbf{publications for full review}};

    \node[startstop, draw, text width=11cm, align=center, right of=total,  xshift=7cm, yshift=-4cm] (abstract excl)
    {\textbf{240 publications excluded:}\\[0.5em]
    \begin{itemize}
        \item 109 removed to include only ``free full text"/``open access" results
        \item 42 duplicate documents removed
        \item 83 publications excluded through screening of titles and abstracts
        \item 6 publications not accessible through institution
    \end{itemize}
    };

   \node[startstop, draw, text width=5cm, align=center, below of=full, yshift=-5cm] (included) 
    {$\boldsymbol{T=41}$ \textbf{publications for inclusion}};

    \node[startstop, draw, text width=11cm, align=center, right of=full, xshift=7cm, yshift=-4cm] (full excl) {\textbf{85 publications excluded after full review:}\\[0.5em] 
        \begin{itemize}
        \item 19 publications did not elicit judgements from subject matter experts
       \item 19 publications did not elicit parameters in the design or analysis stages of clinical trials
        \item 26 publications did not elicit treatment outcome, or borrowing parameters 
        \item 14 publications did not present the method in sufficient detail to allow for comprehension or reproduction
        \item 1 publication described a non-human clinical trial
        \item 2 publications referenced elicitation procedures used in trials that showed up in separate publications in the database search
        \item 4 publications are existing literature/systematic reviews of elicitation methods 
    \end{itemize}
    };

    \node[startstop, draw, text width=5cm, align=center, below of=included] (rare) 
    {$\boldsymbol{T_1=14}$ \textbf{relevant rare disease publications}};
    
    \draw [arrow] (total) -- (full);
    \draw [arrow] (full) -- (included);
    \draw [arrow] (included) -- (rare);
    % Branching arrows to exclusion boxes
    \draw [arrow] ($(total.south)!0.5!(full.north)$) |- (abstract excl.west);
    \draw [arrow] ($(full.south)!0.5!(included.north)$) |- (full excl.west);
    
\end{tikzpicture}
\caption{PRISMA flow diagram} \label{PRISMA}
\end{figure}
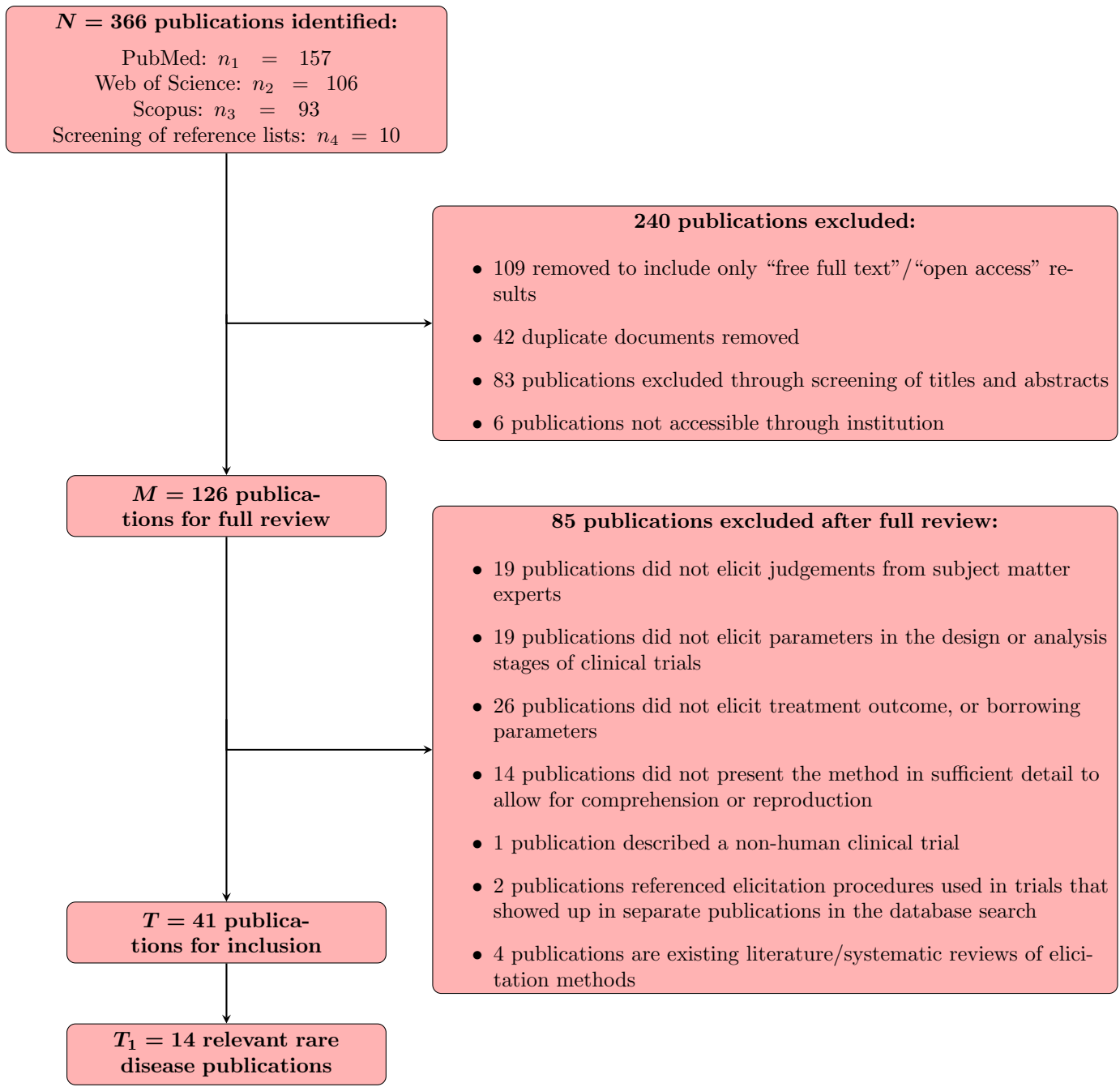

The database search identified 366 publications, of which 251 were free to access through our host institution. Duplicate removal eliminated 42 publications, before a further 83 were removed at the abstract screening stage. This resulted in 126 publications being eligible for full text review, of which 85 were excluded based on the inclusion criteria. The reasons for exclusion were often not mutually exclusive, and so they were applied in the order given in Figure \ref{PRISMA}, which details the first cause for non-eligibility of the excluded publications. %This information is summarised in Figure \ref{PRISMA}.
\\

\begin{figure}
    \centering
    \includegraphics[width=\textwidth]{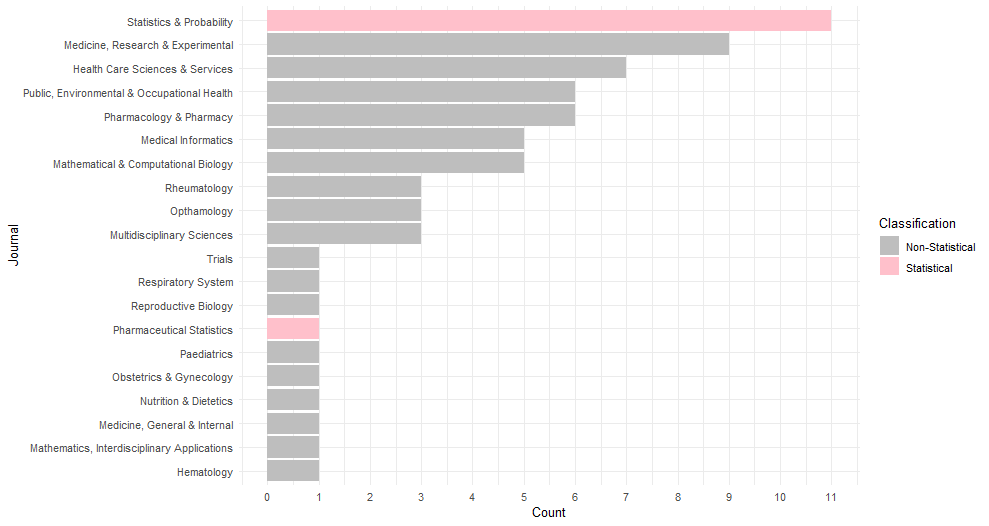}
    \caption{Bar plots showing the frequencies of different journals and their resulting classifications. Note: many journals had multiple classifications.}
    \label{fig: Frequencies of statistical and non-statistical journals}
\end{figure}

\noindent Of the 41 included publications, only 12 were published in statistical journals according to the Journal Citation Reports Classification, as seen in Figure \ref{fig: Frequencies of statistical and non-statistical journals}. The journal types and their frequencies are displayed in Figure \ref{fig: Frequencies of statistical and non-statistical journals} and Table \ref{paper characteristics}. Note that many journals were included in multiple categories. 
\\

\begin{figure}
    \centering
    \includegraphics[width=\textwidth]{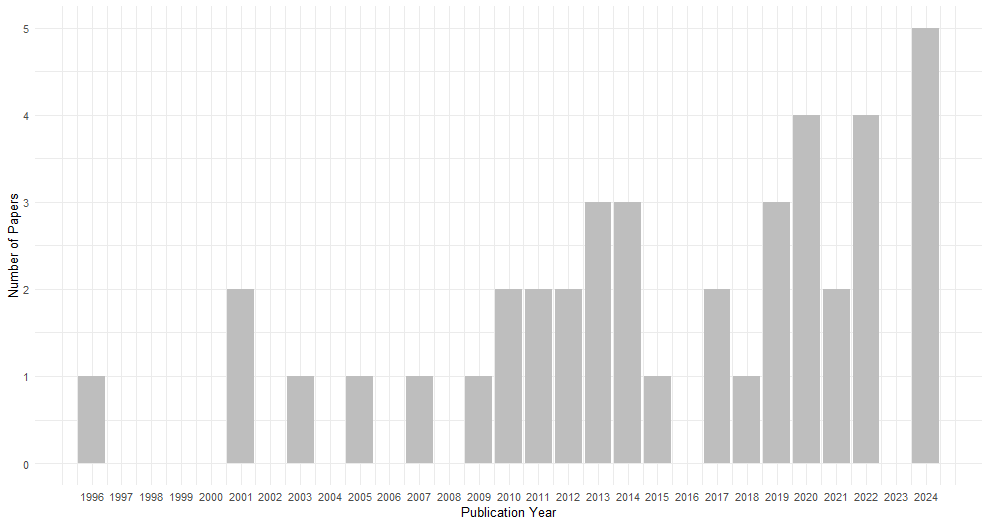}
    \caption{Bar plot showing the year of publication of the included publications}
    \label{year of pub}
\end{figure}

\noindent The included publications span a period of 28 years, from 1996 to (October) 2024, though most were published from 2009 onwards. There is a general upward trend in the number of publications each year, as illustrated in Figure \ref{year of pub}.
\begin{table}
        \centering
        \begin{tabular}{|c|c|} 
        \hline
        \textbf{Journal Citation Report Classification} & \textbf{Number (percentage)} \\
           \hline
           Health Care Sciences \& Services & 7 (17.1\%) \\
           Hematology & 1 (2.4\%) \\
           Mathematical \& Computational Biology & 5 (12.2\%) \\
           Mathematics, Interdisciplinary Applications & 1 (2.4\%) \\
           Medical Informatics & 5 (12.2\%) \\
           Medicine, General \& Internal & 1 (2.4\%) \\
           Medicine, Research \& Experimental & 9 (22.0\%) \\
           Multidisciplinary Sciences & 3 (7.3\%) \\
           Nutrition \& Dietetics & 1 (2.4\%) \\
           Obstetrics \& Gynecology & 1 (2.4\%) \\
           Opthamology & 3 (7.3\%) \\
           Paediatrics & 1 (2.4\%) \\
           Pharmaceutical Statistics & 1 (2.4\%) \\
           Pharmacology \& Pharmacy & 6 (14.6\%) \\
           Public, Environmental \& Occupational Health & 6 (14.6\%) \\
           Reproductive Biology & 1 (2.4\%) \\
           Respiratory System & 1 (2.4\%) \\
           Rheumatology & 3(7.3\%) \\
           Statistics \& Probability & 11 (26.8\%) \\
           Trials & 1 (2.4\%) \\
           \hline
           \textbf{Contribution Type} &\textcolor{white}{placeholder} \\
           \hline
           Applied & 28 (68.3\%) \\
           Methodological/Theoretical & 13 (31.7\%) \\
           \hline
        \end{tabular}
        \caption{Summary of characteristics of included publications. Note: many journals were included in multiple categories of the Journal Citation Report Classification.}
        \label{paper characteristics}
    \end{table} 
\subsection{Methodologic assessment}
\subsubsection{Information and training provided to experts}
Of the 41 included publications, 22 reported training and preparing experts in advance of the elicitation, with $\frac{12}{14}$ rare disease publications reporting training compared to $\frac{10}{27}$ non-rare disease publications. Providing training is the recommendation in elicitation literature in order to ensure the experts share a common understanding of the process, and of relevant statistical concepts, but among the 19 publications not reporting training it was suggested that it was difficult to perform in practice\textsuperscript{\cite{hiance2009practical}}. However, there are a variety of ways to approach this, which we have loosely split into two categories: methodological and context-specific training. 
\\

\noindent Methodological training encompasses any materials or presentations provided to experts prior to the elicitation exercise. These may include general information about the elicitation process\textsuperscript{\cite{jansen2020elicitation}}, foundational concepts from the broader statistical paradigm such as probability or Bayesian methods\textsuperscript{\cite{thallbayesian,dallow2018better,ramanan2019defining,papadopoulou2024elicitation,hampson2015elicitation,maher2024estimating,desai2021prior}}, or specific descriptions of the elicitation task itself, including illustrative guides or aids\textsuperscript{\cite{see2012prior,pokharel2021effectiveness,brown2010effectiveness}}. For example, Kinnersley et al.\textsuperscript{\cite{kinnersley2013structured}} presented relevant terminology to the experts involved in the study, as well as discussing the available published elicitation methods with recommendations for usage, while Holzhauer et al. (2022)\textsuperscript{\cite{holzhauer2022eliciting}} claim that there is no need for the experts to have a thorough understanding of probability or statistical theory. We say that such training is not trial specific, and is generally applicable to any trial, or other elicitation settings. 
\\

\noindent On the other hand, context-specific training is anything provided to experts that is relevant only to that specific trial, condition or drug\textsuperscript{\cite{papadopoulou2024elicitation,maher2024estimating,see2012prior,pokharel2021effectiveness,brown2010effectiveness,hemming2012bayesian,jansen2024uk}}. One example would be providing an evidence or study dossier, which contains relevant publications, current medical evidence, and discussions about the intervention(s) or condition(s) on which the trial is based\textsuperscript{\cite{jansen2020elicitation,dallow2018better,papadopoulou2024elicitation,hampson2015elicitation,maher2024estimating,pokharel2021effectiveness,lan2022remote,hampson2022improving,hampson2014bayesian,cornelius2024treating}}. The design and structure of the trial and associated methods could also be included, as well as a description of how and why the prior elicitation and resulting distribution will be used\textsuperscript{\cite{white2007eliciting}}. Some publications also used sample questionnaires, hypothetical trials, or elicited general knowledge quantities, so that experts were able to complete a `dummy elicitation'\textsuperscript{\cite{jansen2020elicitation,thallbayesian,kinnersley2013structured}}. Allowing experts to complete a ‘no-pressure’ example is beneficial; it not only aids their understanding of the process, but also ensures they engage with the material, as it is often facilitated in a format similar to the actual elicitation. However, it can be difficult to come up with relevant examples and scenarios to use, particularly in rare disease trials. 
\\ 

\noindent In practice, combinations of the above can be used, with 9 publications\textsuperscript{\cite{dallow2018better,ramanan2019defining,papadopoulou2024elicitation,hampson2015elicitation,maher2024estimating,,see2012prior,pokharel2021effectiveness,brown2010effectiveness,jansen2024uk}} reporting on multiple training types, such as a worked example or sample questionnaire, alongside relevant statistical training or a summary of current evidence. Crucially, it is the delivery of these training materials that ensures effective dissemination amongst experts; some materials, particularly evidence dossiers, are provided in advance of the elicitation process, but it is impossible to police whether experts have actually engaged adequately, or at all, with them. To combat this, if the subsequent elicitation method involves a face-to-face meeting, it is not uncommon for this to be reviewed prior to the elicitation exercise. There are four main benefits of this: 
\begin{itemize}
    \item it reminds (or introduces) the experts of the required context; 
    \item it provides opportunity to ask for clarification; 
    \item it makes the facilitators aware of any additional studies the experts believe should have been included in the preparation; 
    \item it allows the experts to identify and declare any conflicts of interest. 
\end{itemize} 
\noindent Holding a pre-trial planning meeting\textsuperscript{\cite{thallbayesian}}, creating an e-learning module that experts can complete in their own time\textsuperscript{\cite{dallow2018better,holzhauer2022eliciting}}, or providing a sample questionnaire/example elicitation\textsuperscript{\cite{jansen2020elicitation,ramanan2019defining,papadopoulou2024elicitation,hampson2015elicitation,holzhauer2022eliciting,jansen2024uk,cornelius2024treating,johnson2010valid,johnson2011effect,diamond2014expert}}, also provide training in which experts can be confirmed to have engaged, but such methods bring their own criticisms in the additional time commitment required by experts. 
\\

\noindent Providing training containing results from previous studies has been criticised in the literature, with some authors claiming that providing such information to experts ahead of the elicitation can shift their mindsets, and make them feel as though the elicitation is a `test' in which they have to `match' the evidence provided in the study dossier, since this would introduce bias into the study as the experts would be less likely to provide their actual beliefs\textsuperscript{\cite{pokharel2021effectiveness}}. Pokharel et al. (2021)\textsuperscript{\cite{pokharel2021effectiveness}} address this by highlighting potential biases to experts, clarifying that the goal of the study is to understand their opinions, and pointing out that there are many reasons why their beliefs may not match the provided evidence. Brown et al. (2010)\textsuperscript{\cite{brown2010effectiveness}} disagree, noting the value of provision of an example or training exercise for reducing bias.  
\\

\noindent Regardless of concerns about the introduction of bias and additional time demands, training should be an essential part of the elicitation process\textsuperscript{\cite{dallow2018better}}, as it reduces the risk of misconceptions, and gives the experts a better all-round understanding, ensuring they consider all available evidence rather than only evidence of which they are already aware. This could be particularly relevant in a rare disease context, in which experts have seen fewer patients and therefore may be less informed on how a patient population could be related to another, similar, population, or how a general patient might react to an intervention, for example. Therefore, providing an evidence dossier will help them to express their judgements more clearly, and to appreciate the importance of the procedure. This would help to make elicitation a more intuitive, and smoother, process for the experts, potentially making them more likely to engage in future exercises. Moreover, providing training would ensure all experts have an equal knowledge about the elicitation process, allowing for a less strict criteria on their selection since pre-requisite statistical knowledge becomes less relevant. Both of these outcomes could be especially helpful in the rare disease setting where there are likely fewer experts eligible to be selected. 

\subsubsection{Methods and conduct of elicitation}
There was no discernible difference in the methods and conduct of elicitation across all trials and rare disease trials specifically, and, in both, over half of included publications used some variation of the SHeffield ELicitation Framework (SHELF)\textsuperscript{\cite{o2013shelf}}. 
\\

\noindent Aligning with this, it was generally agreed that one-on-one, face-to-face or telephone interviews were preferable to non-synchronous elicitation, since they allow experts to seek clarification and ask questions, hence leading to responses that are more valid and reliable\textsuperscript{\cite{johnson2011effect}}. Most publications made sure to facilitate an in-person elicitation\textsuperscript{\cite{papadopoulou2024elicitation,holzhauer2022eliciting,hampson2014bayesian,johnson2011effect,diamond2014expert,turner2022borrowing,mason2017development,thirard2020integrating,white2005eliciting}}, even if this meant staging it at, for instance, a conference that all included experts would be attending\textsuperscript{\cite{hiance2009practical}}, mitigating the problem that in-person elicitation can be difficult and expensive to organise\textsuperscript{\cite{jansen2024uk}}, particularly when working with rare diseases when experts may have to be selected from a larger geographical area due to the spread of patients, and international differences in management\textsuperscript{\cite{lan2022remote}}. Additionally, facilitating elicitation meetings is time-consuming, both for the trialist and the experts, which was the main justification for publications using other means such as online non-synchronous responses, post, or email, although asynchronous elicitation exercises often have low response and engagement rates. Additionally, they only allow for limited assistance during the elicitation session, and experts are less able to discuss and calibrate their beliefs\textsuperscript{\cite{lan2022remote}}. One publication gave experts a choice of how they would conduct the elicitation (either through an in-person structured interview or online)\textsuperscript{\cite{hemming2012bayesian}}. We would not recommend this as it creates a discrepancy in the reliability of judgements that are elicited, rather than being consistent across all experts. Some publications conducting elicitation exercises in 2020 and 2021 were forced to conduct online elicitations due to the Covid-19 pandemic, but it was unclear whether these included video calls, or other online methods. Using video calls for elicitation could be beneficial when face-to-face elicitation isn't possible, and would help to solve the problems associated with the geographical spread of experts more generally.
\\

\noindent Many publications expressed a preference for using questionnaires\textsuperscript{\cite{ramanan2019defining,papadopoulou2024elicitation,hampson2015elicitation,hemming2012bayesian,mason2017development,browne2017bayesian,mason2020framework,tan2003elicitation,chaloner2001quantifying}}, as recommended in the STEER protocol\textsuperscript{\cite{SteerYork}}. However, these are typically less reliable than structured interviews, since it is not uncommon for experts to give their judgements on the wrong scale when the completion of the questionnaire is not facilitated by a statistician. White et al. (2005)\textsuperscript{\cite{white2005eliciting}} highlight the importance of a face-to-face elicitation, stating that they enable the elicitation of quantiles (rather than constraining the judgements of experts into pre-specified response categories), and for feedback to be given, though they note the practical advantages of written questionnaires. As part of a questionnaire, experts were commonly asked to assign `chips' to `bins' to form a histogram that reflects the mean and uncertainty of their judgements\textsuperscript{\cite{thallbayesian,pokharel2021effectiveness,kinnersley2013structured,holzhauer2022eliciting,hampson2022improving,cornelius2024treating,white2007eliciting,johnson2010valid,johnson2011effect,diamond2014expert,turner2022borrowing,white2005eliciting,abrams1996bayesian,ren2014assurance,sun2013expert,vail2001prospective}}. This is aptly called the `bins and chips', `roulette' or `histogram' method. This method is valued as it is seen as an intuitive way for experts to express their judgements\textsuperscript{\cite{jansen2020elicitation}}. 
\\

\noindent A variation of this approach involves asking experts to draw a distribution that represents their judgements\textsuperscript{\cite{brown2010effectiveness,brownstein2019role}}, but most studies employing visual methods tend to prefer the roulette method. Nonetheless, there are variations that can be interactive and iterative: another variation allows experts to revise their distribution, either through real-time graphical feedback or by having a statistical facilitator translate their judgements into a visual distribution\textsuperscript{\cite{papadopoulou2024elicitation,see2012prior,mason2020framework}}. This can involve multiple revisions until the expert feels their beliefs are adequately captured. Such an approach can be particularly beneficial, as it allows for a more accurate representation of the experts' judgements.
\\

\noindent Some publications choose to feedback to experts in a more explicitly mathematical way; in the rare disease setting, Papadopoulou et al.\textsuperscript{\cite{papadopoulou2024elicitation}} suggest calculating an effective sample size of the expert's prior, which allows the demonstration of outcomes in hypothetical datasets and hence gives opportunity for experts to assess their judgements. They still provided real-time graphical feedback which allowed for an iterative approach. 
\\

\noindent An alternative to the roulette method that is included in the SHELF recommendations and used in some of the included publications is the elicitation of quantiles\textsuperscript{\cite{brown2010effectiveness,thirard2020integrating}}. Again, this is largely implemented through questionnaires and/or interviews. This method has the advantage that quantiles have clear operational definitions, and so the facilitator can guide the expert through a structured process, ensuring that the expert understands the tasks they are completing. This approach is also less susceptible to the influence of choices made by the facilitator, such as the range of the variable considered in the roulette method, and the number and placement of the bins.
\\

\noindent Although the elicitation of the two types of parameters we focus on in this study - treatment outcome and borrowing parameters - was largely similar, Rietbergen et al. (2011)\textsuperscript{\cite{rietbergen2011incorporation}} adopted a novel approach to assessing similarity between historical studies and the current trial. Rather than directly eliciting parameter values or distributions, they asked an expert to rank previous studies based on their relevance to the current trial. 
\\

\begin{figure}
    \centering
    \includegraphics[width=\textwidth]{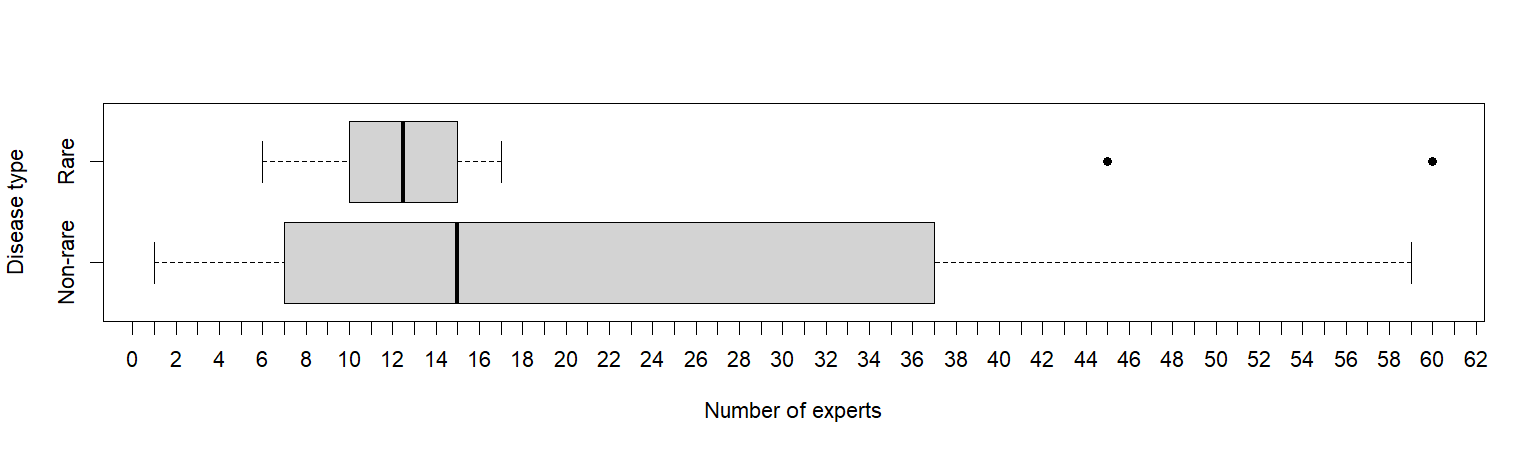}
    \caption{Boxplots showing the number of experts involved in the elicitation of included trials across the two disease types}
    \label{experts}
\end{figure}

\noindent Overall, there was significant variation in the number of experts included across studies, even within a relatively small sample, as illustrated in Figure \ref{experts}. Johnson et al. (2011)\textsuperscript{\cite{johnson2011effect}} noted the lack of consensus on the appropriate sample size for elicitation studies and proposed using the Central Limit Theorem to determine the number of experts required to approximate a normal distribution for the mean of the group’s beliefs. This is a unique approach, in a pool of publications in which little or no explanation was provided for the number of experts involved in the study. There is evidence to suggest that 4-10 experts is recommended in most commonly used elicitation protocols\textsuperscript{\cite{Quigley2018}}, since more experts become difficult to manage, potentially leading to lower quality elicitations from individual experts. Hakkma et al.(2014)\textsuperscript{\cite{haakma2014belief}} argued that beyond 12 experts, the marginal benefit for including more experts decreases, though this study is not directly related to trials and highlights how elicitation in trial contexts seem to be less structured than in health economic and cost-effectiveness contexts.
\\

\noindent Regardless of the method used, very few publications describe how they obtain fitted parametric distributions after the elicitation is complete, unless the distribution is pre-specified.  

\subsubsection{Decomposition of variables and resulting distributions}
Although our study focuses on treatment and borrowing parameters, these can be elicited through the decomposition or transformation of other variables to make the quantities more accessible and understandable to experts. For example, parameters such as log-odds ratios are often preferred over direct probability differences because they better model judgements about a bounded quantity within a normal distribution\textsuperscript{\cite{hampson2014bayesian}}. However, it may be easier for clinicians to provide probability differences rather than odds ratios. While it may be more straightforward to ask questions about the latter, this could conflict with clinician preferences, suggesting that questions and parameters might need to be adapted accordingly\textsuperscript{\cite{hampson2014bayesian}}. For example, difficulties can arise when interaction parameters are elicited, since they are not observable in practice, and therefore cannot be elicited directly. White et al. (2005)\textsuperscript{\cite{white2005eliciting}} suggest overcoming this by eliciting a plausible value of the treatment effect, and the prior probability of a qualitative interaction.  
\\

\noindent An intuitive way for experts to express their judgements is through proportions, as this most closely resembles clinical decision-making\textsuperscript{\cite{lan2022remote}}. Alhussain et al. (2020)\textsuperscript{\cite{alhussain2020assurance}} elicited lower and upper bounds, which were interpreted as the 5th and 95th percentiles on the proportion of patients who have outcomes in a given interval. This allowed for comparison between the treatment and control arms - essentially asking experts ``by how much is this intervention better than the control?" through upper and lower limits on the proportion of patients that would respond well to the interventions. Experts also had little difficulty expressing their beliefs in terms of probability of improvement using treatment over control, and in giving mean differences in response rates between the two treatments\textsuperscript{\cite{white2005eliciting}}. Hazard ratios and hazard rates, as well as summary statistics like mean and mode, were also deemed to be observable quantities that were understandable enough to be elicited from experts\textsuperscript{\cite{hiance2009practical,ren2014assurance,moatti2013modeling}}, which follows the general consensus in the literature that experts should only be asked to assess distributions of observable quantities\textsuperscript{\cite{white2005eliciting}}. Salsbury et al. (2024)\textsuperscript{\cite{salsbury2024assurance}} reiterate this, while still noting the possible need to adjust quantities. They state that while survival probability at a time $t$, the median survival time on an experimental treatment, and the size of difference between control and experimental survival curves are considered observable quantities on a treatment effect, experts prefer to make judgements about hazard ratios at time $t$, and the maximum hazard ratio and when it occurs. 
\\ 

\noindent Normal distributions are often familiar to experts, and so sometimes the Normal distribution parameters were directly elicited\textsuperscript{\cite{abrams1996bayesian,alhussain2020assurance}}, although this became more difficult when the resulting distribution was asymmetric. Imposing parametric distributions, particularly symmetric ones like the Normal distribution, can be problematic as it forces experts to adopt symmetric beliefs. This may also cause confusion, as experts might not realise they have the option to express asymmetric beliefs, with some experts commonly feeling that a symmetric distribution is ``required", since they have continually been presented with Normal distributions during statistical training\textsuperscript{\cite{dallow2018better}}. For example, the asymmetry of the beta distribution, makes it more difficult for experts to intuitively select a mean and standard deviation\textsuperscript{\cite{wu2008elicitation}}. Gaithwate et al. (2005)\textsuperscript{\cite{garthwaite2005statistical}} have provided evidence to suggest that experts are not good at specifying standard deviations directly, since this is not an observable quantity, and experts have consistently stated that they find it more intuitive to provide most likely values rather than values at the tails of a distribution\textsuperscript{\cite{kinnersley2013structured}}, so steps must be taken to ensure the experts can adequately express their full judgements.
\\

\noindent Only 8 of the 41 included publications did not result in a parametric distribution, instead using a nonparametric alternative. Examples of the resulting parametric distributions are Normal (including bi/multivariate Normal, mixture of Normals, and split Normal), log-normal, beta, binomial, gamma, Bayesian Cox proportional hazards models, and Weibull survival models. Only one publication reported a comparison of a few distributions\textsuperscript{\cite{pokharel2021effectiveness}}, despite this being one of the recommendations made in SHELF. In cases where distributions show a lack of fit, especially pre-specified parametric distributions, it appears that they were often kept for mathematical convenience.
\\

\noindent Overall, there was little difference in terms of the elicited parameters and the overall resulting distributions between the rare and non-rare disease settings. 

\subsubsection{Aggregation methods}
The publications in the study revealed ten distinct methods for aggregating expert judgements into a single prior. Of those including the judgements of more than one expert, only one publication did not aggregate expert judgements - a probability distribution on the parameters was not specified\textsuperscript{\cite{chaloner2001quantifying}}. There were two methods used overwhelmingly amongst the included studies: behavioural aggregation\textsuperscript{\cite{jansen2020elicitation,ramanan2019defining,hampson2015elicitation,maher2024estimating,desai2021prior,kinnersley2013structured,holzhauer2022eliciting,jansen2024uk,hampson2014bayesian,cornelius2024treating,johnson2011effect,ren2014assurance,salsbury2024assurance}}, and mathematical pooling\textsuperscript{\cite{thallbayesian,see2012prior,hampson2022improving,white2007eliciting,mason2017development,thirard2020integrating,white2005eliciting,mason2020framework,abrams1996bayesian,brownstein2019role}}, with 12 and 17 publications using these approaches, respectively. 
\\

\noindent Behavioural aggregation is the recommendation for aggregating expert judgements in SHELF\textsuperscript{\cite{o2013shelf}} and requires experts to agree a consensus prior through a guided discussion with a statistical facilitator. They should produce a single prior distribution that they collectively agree a `rational impartial observer' would determine upon observing the discussions\textsuperscript{\cite{dallow2018better}}. This method has advantages over its mathematical counterparts in that it always results in distributions that have sensible practical interpretations\textsuperscript{\cite{williams2021comparison}}. However, there is a risk of overconfident group members and strong personalities dominating the discussion\textsuperscript{\cite{ramanan2019defining}}, leading to a prior distribution that does not fully incorporate every expert's beliefs, and that is biased towards one or a few experts. On the other hand, sometimes experts may be unable to reach a consensus, and may instead split into groups with other experts who share their beliefs, resulting in multiple distributions, which would then require other methods to be employed to aggregate\textsuperscript{\cite{williams2021comparison}}. The notion of the variation in expert judgements being too great to allow for efficient behavioural aggregation is particularly prevalent in the rare disease setting, in which the small patient population limits the experience of experts and may mean that experts see patients exhibiting different symptoms, and with different characteristics. Despite this, behavioural aggregation was still the prevailing method in the rare disease publications, with $\frac{8}{14}$ publications reporting this as their method of aggregation. This means that it was more common in this setting than in the non-rare setting. Obtaining a prior with sensible practical interpretations cannot be overlooked, and when the method does result in a final consensus prior it can be highly effective, and some degree of interaction between experts is important in ironing our misconceptions and uncertainties. Similarly, if experts come from different specialisms, behavioural aggregation can be extremely useful in providing experts with the platform to discuss and consider different views and perspectives to inform the consensus prior distribution, rather than mathematically weighting them\textsuperscript{\cite{hampson2014bayesian}}.
\\

\noindent The other prominent method was mathematical pooling, which forms a prior by taking the weighed arithmetic (linear pooling) or geometric mean (log linear pooling) of each expert's prior distributions. Linear pooling is often preferred, since log linear pooling can cause problems in the case where there is little similarity or overlap between individual expert priors, resulting in high density between expert priors\textsuperscript{\cite{williams2021comparison}}. This could be particularly problematic in the rare disease setting, where we expect that there may be more variation between individual expert priors. The main appeal of using linear pooling is that it can reduce biases introduced by overconfidence. It is also argued to be easier to implement\textsuperscript{\cite{lan2022remote}}. 
\\

\noindent Mathematical pooling was most commonly done by equally weighting experts\textsuperscript{\cite{thallbayesian,see2012prior,jansen2024uk,hampson2022improving,white2007eliciting,mason2017development,thirard2020integrating,white2005eliciting,mason2020framework,abrams1996bayesian,brownstein2019role}}, as recommended in the STEER protocol\textsuperscript{\cite{SteerYork}}, mainly because it can be difficult, and sensitive, to quantify who is a `better' expert, particularly with rare diseases where there is less external knowledge and evidence to inform this decision. Empirical evidence suggests that weighting experts in this way leads to a well-calibrated distribution, but one that is less informative since it includes more extreme judgements. When applying this method in rare disease trials, this approach could be problematic since the small patient sample sizes in the trials requires the prior to be as informative as possible to supplement the limited trial data. Conversely, constructing a prior with unequally weighted experts leads to a prior which can be more informative, so this could perhaps be more applicable to the rare disease setting, though the question remains of how to establish the weights. 
\\

\noindent Hampson et al. (2015)\textsuperscript{\cite{hampson2015elicitation}} discuss involving a second set of experts who decide how to weight the competing judgements of the experts in the initial elicitation, while other publications try to incorporate `seed' questions in the original questionnaire\textsuperscript{\cite{papadopoulou2024elicitation}}. The aim of those is to `test' the experts' knowledge of the area by asking them to provide information on known quantities. However, it can be difficult to come up with questions that relate closely enough to the trial to be sufficient to assess the experts, but that are different enough to not already be related to elicitation quantities\textsuperscript{\cite{mason2020framework}}. This problem is exacerbated in rare diseases since less is known about the condition and therefore any associated quantities. This latter method is known as the `Cooke'/`classical' method \textsuperscript{\cite{Quigley2018}}, and typically results in informative distributions, but involves more work for both the experts and the statisticians. It also requires assumptions about expert performance on the seed and elicitation questions being highly correlated.
\\

\noindent Other aggregation methods in the included publications are simple averaging of means and standard deviations\textsuperscript{\cite{hemming2012bayesian}}, or using the mean probability values to construct an overall histogram\textsuperscript{\cite{tan2003elicitation}}. Brown et al. (2010) \textsuperscript{\cite{brown2010effectiveness}} discuss summing the heights of the bars in each expert-elicited prior and weighting them based on the total sum of heights. Using the median of the elicited values allows the pooled data to represent the judgement of a `typical' expert and is not influenced by extreme opinions\textsuperscript{\cite{turner2022borrowing}}, although it is unclear how this incorporates uncertainty. Each of these approaches could be described as ``ad hoc" and do not follow good practice in the elicitation literature.
\\

\noindent When expert judgements vary greatly, it has been suggested to be beneficial to split the group into optimistic and pessimistic subgroups\textsuperscript{\cite{pokharel2021effectiveness}}. This is similar to what was discussed in the behavioural case, except that instead of the experts splitting themselves, the process is mathematical and splits the experts based on the elicited individual quantiles. Although this should result in two informative priors, there is a problem in that there is more than one prior, which could result in conflicting conclusions if they are used in separate Bayesian analyses.  
\\

\noindent Diamond et al. (2014)\textsuperscript{\cite{diamond2014expert}} take a novel approach involving simulation: they propose sampling values sequentially from each participant's prior 100,000 times for each treatment arm and calculating a relative risk with each repetition. These relative risk distributions can be averaged to obtain a participant-wide distribution. However, we can still foresee this resulting in non-informative distributions if expert judgements differ greatly, as well as the computational intensity of doing this procedure if a large number of experts are used. This is a variation on equally weighting the experts' priors.

\section{Discussion}
Overall, the methodology used for elicitation in rare and non-rare diseases appears largely similar; there is no recommendation or framework for adapting standard elicitation methods used in clinical trials to the rare disease setting, despite the unique challenges in the latter. The absence of tailored elicitation approaches represents an important opportunity for methodological development, particularly given the reliance on expert judgement in the design and analysis of rare disease trials. That said, it is difficult to say for sure, since reporting practices in both areas are inconsistent, with little to no justification for expert selection in most cases, despite the availability of SHELF templates. This suggests a gap in understanding of elicitation methodology amongst those carrying it out in the clinical trial setting.
\\

\noindent Many elicitations carried out used a questionnaire, which can be effective when completed under the guidance of a statistical facilitator. However, often, this is not the approach taken, with some or all experts completing it remotely. In our experience conducting elicitations, this is an undesirable approach, as it means that experts are less likely to seek clarification, and it is difficult to ensure that the experts have completed the required training, and read and understood the evidence dossier. This can lead to major discrepancies between experts, due to individual experts misinterpreting elicitation variables, evidence or elicitation tasks, resulting in, for example, quantities being elicited on a completely different scale. That said, many remote elicitations were enforced due to the Covid-19 pandemic, rather than being the trialist's preference, although video calls could be routinely employed as a natural alternative in these circumstances. Taking this approach would be particularly useful in rare diseases even beyond the pandemic, since the geographical spread of experts can make a face-to-face approach more difficult.
\\

\noindent There is an overall agreement that individual expert judgements should be aggregated into a single prior distribution. In many cases this was to facilitate a Bayesian analysis upon receipt of data at trial interim analysis or at the end of the trial. Emphasis was placed on the importance of expert interaction - even with mathematical aggregation it is possible for experts to discuss, and perhaps to have the opportunity to revise their own prior distribution as a result of this discussion\textsuperscript{\cite{IDEA}}. 
\\

\noindent Only a minority of studies considered repeating the elicitation with a second, distinct group to assess between-expert group variability empirically. This was deemed too resource-intensive, especially in rare disease contexts where the pool of eligible experts is small. Many studies acknowledged the challenge of balancing accuracy with minimising the demands and burden on experts.
\\

\noindent A potential limitation of this study is the number of databases searched. The choice was made based on the authors' knowledge of publications in this field, and limited based on practicality of timely screening. Therefore, an extension of this study could include searching alternative databases. This study is scheduled to be reviewed and updated in 2027.
\\

\noindent Furthermore, this study has limited consideration of the control of heuristics and biases, and post elicitation procedures (such as feedback to experts after having observed data). This was decided in the interest of time and practicality, but can also be justified by the idea that these are unlikely to vary substantially between the two disease settings involved in the comparison. Moreover, the focus of the review was on the statistical conduct, and commonality, of the elicitation methods, and less on the evaluation of their overall performance. Bojke et al.\textsuperscript{\cite{bojke2021developing}} discuss how such concerns are treated across elicitation protocols for Health Technology Assessments, and including them in this study would be a relatively straightforward extension of our data extraction template. Such considerations of biases and vested interests are integral to the SHELF procedure\textsuperscript{\cite{dallow2018better}}, so consideration of these aspects was assumed in most publications, since that was the prevalent framework used. Kinnersley et al. (2013)\textsuperscript{\cite{kinnersley2013structured}} claim that when expert judgements are used to inform the prior, the clinical trial literature provides little insight into the feasibility and reliability of eliciting such beliefs. Their comment serves to emphasise how any recommended procedures and protocols that exist in the trial setting are often adapted or ignored. 

\section*{Declarations}
\subsection*{Ethics approval and consent to participate}
This study does not involve any animals, humans, human data, human tissue or plants, so ethical approval is not applicable. 
\subsection*{Consent for publication}
This study does not contain data from any individual person, so consent for publication is not applicable. 
\subsection*{Availability of data and material}
This study does not contain any data, and the publications discussed are available on named public databases, so availability of data and material is not applicable. 
\subsection*{Competing interests}
The authors declare that they have no competing interests.
\subsection*{Funding}
This study was completed as part of the requirements of MRC-NIHR Trials and Methodology Research Partnership Doctoral Training Partnership (TMRP DTP) PhD qualification, though no direct funding was received for completion of this manuscript. 
\subsection*{Authors' contributions}
All authors contributed to the design of the study. LMM screened all  publications, extracted all relevant data, and drafted the manuscript. JMSW, KJW, and NW each reviewed a subset of included papers to ensure consistency in applying the inclusion and exclusion criteria. All authors contributed to the interpretation of the findings, and read and approved the final manuscript.
\subsection*{Acknowledgements}
Not applicable. 
\printbibliography

@article{stefan2022expert,
  title={Expert agreement in prior elicitation and its effects on Bayesian inference},
  author={Stefan, Angelika M and Katsimpokis, Dimitris and Gronau, Quentin F and Wagenmakers, Eric-Jan},
  journal={Psychonomic Bulletin \& Review},
  volume={29},
  number={5},
  pages={1776--1794},
  year={2022},
  publisher={Springer}
}

@article{johnson2010methods,
  title={Methods to elicit beliefs for Bayesian priors: a systematic review},
  author={Johnson, Sindhu R and Tomlinson, George A and Hawker, Gillian A and Granton, John T and Feldman, Brian M},
  journal={Journal of clinical epidemiology},
  volume={63},
  number={4},
  pages={355--369},
  year={2010},
  publisher={Elsevier}
}

@article{mikkola2024prior,
  title={Prior knowledge elicitation: The past, present, and future},
  author={Mikkola, Petrus and Martin, Osvaldo A and Chandramouli, Suyog and Hartmann, Marcelo and Abril Pla, Oriol and Thomas, Owen and Pesonen, Henri and Corander, Jukka and Vehtari, Aki and Kaski, Samuel and others},
  journal={Bayesian Analysis},
  volume={19},
  number={4},
  pages={1129--1161},
  year={2024},
  publisher={International Society for Bayesian Analysis}
}

@article{o2013shelf,
  title={SHELF: the Sheffield elicitation framework},
  author={O’Hagan, Anthony and Oakley, J},
  journal={Reference Source},
  year={2013}
}

@article{jansen2020elicitation,
  title={Elicitation of prior probability distributions for a proposed Bayesian randomized clinical trial of whole blood for trauma resuscitation},
  author={Jansen, Jan O and Wang, Henry and Holcomb, John B and Harvin, John A and Richman, Joshua and Avritscher, Elenir and Stephens, Shannon W and Truong, Van Thi Thanh and Marques, Marisa B and DeSantis, Stacia M and others},
  journal={Transfusion},
  volume={60},
  number={3},
  pages={498--506},
  year={2020},
  publisher={Wiley Online Library}
}

@article{kinnersley2013structured,
  title={Structured approach to the elicitation of expert beliefs for a Bayesian-designed clinical trial: a case study},
  author={Kinnersley, Nelson and Day, Simon},
  journal={Pharmaceutical statistics},
  volume={12},
  number={2},
  pages={104--113},
  year={2013},
  publisher={Wiley Online Library}
}

@article{lan2022remote,
  title={Remote, real-time expert elicitation to determine the prior probability distribution for Bayesian sample size determination in international randomised controlled trials: Bronchiolitis in Infants Placebo Versus Epinephrine and Dexamethasone (BIPED) study},
  author={Lan, Jingxian and Plint, Amy C and Dalziel, Stuart R and Klassen, Terry P and Offringa, Martin and Heath, Anna and Pediatric Emergency Research Canada (PERC) KIDSCAN/PREDICT BIPED Study Group},
  journal={Trials},
  volume={23},
  number={1},
  pages={279},
  year={2022},
  publisher={Springer}
}

@article{maher2024estimating,
  title={Estimating the effect of nintedanib on forced vital capacity in children and adolescents with fibrosing interstitial lung disease using a bayesian dynamic borrowing approach},
  author={Maher, Toby M and Brown, Kevin K and Cunningham, Steven and DeBoer, Emily M and Deterding, Robin and Fiorino, Elizabeth K and Griese, Matthias and Schwerk, Nicolaus and Warburton, David and Young, Lisa R and others},
  journal={Pediatric Pulmonology},
  volume={59},
  number={4},
  pages={1038--1046},
  year={2024},
  publisher={Wiley Online Library}
}

@article{white2007eliciting,
  title={Eliciting and using expert opinions about dropout bias in randomized controlled trials},
  author={White, Ian R and Carpenter, James and Evans, Stephen and Schroter, Sara},
  journal={Clinical Trials},
  volume={4},
  number={2},
  pages={125--139},
  year={2007},
  publisher={Sage Publications Sage UK: London, England}
}

@misc{thallbayesian,
  title={Bayesian treatment comparison using parametric mixture priors computed from elicited histograms. 2019; 28: 404-18},
  author={Thall, PF and Ursino, M and Baudouin, V and Alberti, C and Zohar, S}
}

@article{dallow2018better,
  title={Better decision making in drug development through adoption of formal prior elicitation},
  author={Dallow, Nigel and Best, Nicky and Montague, Timothy H},
  journal={Pharmaceutical Statistics},
  volume={17},
  number={4},
  pages={301--316},
  year={2018},
  publisher={Wiley Online Library}
}

@article{johnson2011effect,
  title={Effect of warfarin on survival in scleroderma-associated pulmonary arterial hypertension (SSc-PAH) and idiopathic PAH. Belief elicitation for Bayesian priors},
  author={Johnson, Sindhu R and Granton, John T and Tomlinson, George A and Grosbein, Haddas A and Hawker, Gillian A and Feldman, Brian M},
  journal={The Journal of rheumatology},
  volume={38},
  number={3},
  pages={462--469},
  year={2011},
  publisher={The Journal of Rheumatology}
}

@misc{SteerYork,
  title = {STEER: Economic Evaluation},
  url = {https://www.york.ac.uk/che/economic-evaluation/steer/},
}

@article{papadopoulou2024elicitation,
  title={Elicitation of expert prior opinion to design the BARJDM trial in juvenile dermatomyositis},
  author={Papadopoulou, Charalampia and Martin, Neil and Rafiq, Nadia and McCann, Liza and Varner, Giulia and Nott, Kerstin and Compeyrot-Lacassagne, Sandrine and Leandro, Maria and Foley, Charlene and Warrier, Kishore and others},
  journal={Rheumatology},
  volume={63},
  number={12},
  pages={3271--3278},
  year={2024},
  publisher={Oxford University Press}
}

@article{hampson2014bayesian,
  title={Bayesian methods for the design and interpretation of clinical trials in very rare diseases},
  author={Hampson, Lisa V and Whitehead, John and Eleftheriou, Despina and Brogan, Paul},
  journal={Statistics in medicine},
  volume={33},
  number={24},
  pages={4186--4201},
  year={2014},
  publisher={Wiley Online Library}
}

@article{wu2008elicitation,
  title={Elicitation of a beta prior for Bayesian inference in clinical trials},
  author={Wu, Yujun and Shih, Weichung J and Moore, Dirk F},
  journal={Biometrical Journal: Journal of Mathematical Methods in Biosciences},
  volume={50},
  number={2},
  pages={212--223},
  year={2008},
  publisher={Wiley Online Library}
}

@article{garthwaite2005statistical,
  title={Statistical methods for eliciting probability distributions},
  author={Garthwaite, Paul H and Kadane, Joseph B and O'Hagan, Anthony},
  journal={Journal of the American statistical Association},
  volume={100},
  number={470},
  pages={680--701},
  year={2005},
  publisher={Taylor \& Francis}
}

@article{pokharel2021effectiveness,
  title={Effectiveness of initial methotrexate-based treatment approaches in early rheumatoid arthritis: An elicitation of rheumatologists’ beliefs}, 
  author={Pokharel, Gyanendra and Deardon, Rob and Johnson, Sindhu R and Tomlinson, George and Hull, Pauline M and Hazlewood, Glen S},
  journal={Rheumatology},
  volume={60},
  number={8},
  pages={3570--3578},
  year={2021},
  publisher={Oxford University Press}
}

@article{chaloner2001quantifying,
  title={Quantifying and documenting prior beliefs in clinical trials},
  author={Chaloner, Kathryn and Rhame, Frank S},
  journal={Statistics in medicine},
  volume={20},
  number={4},
  pages={581--600},
  year={2001},
  publisher={Wiley Online Library}
}

@article{williams2021comparison,
  title={A comparison of prior elicitation aggregation using the classical method and SHELF},
  author={Williams, Cameron J and Wilson, Kevin J and Wilson, Nina},
  journal={Journal of the Royal Statistical Society Series A: Statistics in Society},
  volume={184},
  number={3},
  pages={920--940},
  year={2021},
  publisher={Oxford University Press}
}

@article{ramanan2019defining,
  title={Defining consensus opinion to develop randomised controlled trials in rare diseases using Bayesian design: An example of a proposed trial of adalimumab versus pamidronate for children with CNO/CRMO},
  author={Ramanan, AV and Hampson, LV and Lythgoe, H and Jones, AP and Hardwick, B and Hind, H and Jacobs, B and Vasileiou, D and Wadsworth, I and Ambrose, N and others},
  journal={PLoS One},
  volume={14},
  number={6},
  pages={e0215739},
  year={2019},
  publisher={Public Library of Science San Francisco, CA USA}
}

@article{hampson2015elicitation,
  title={Elicitation of expert prior opinion: application to the MYPAN trial in childhood polyarteritis nodosa},
  author={Hampson, Lisa V and Whitehead, John and Eleftheriou, Despina and Tudur-Smith, Catrin and Jones, Rachel and Jayne, David and Hickey, Helen and Beresford, Michael W and Bracaglia, Claudia and Caldas, Afonso and others},
  journal={PLoS One},
  volume={10},
  number={3},
  pages={e0120981},
  year={2015},
  publisher={Public Library of Science San Francisco, CA USA}
}

@article{Quigley2018,
author="Quigley, John
and Colson, Abigail
and Aspinall, Willy
and Cooke, Roger M.",
editor="Dias, Luis C.
and Morton, Alec
and Quigley, John",
title="Elicitation in the Classical Model",
bookTitle="Elicitation: The Science and Art of Structuring Judgement",
year="2018",
publisher="Springer International Publishing",
address="Cham",
pages="15--36",
isbn="978-3-319-65052-4",
doi="10.1007/978-3-319-65052-4_2",
url="https://doi.org/10.1007/978-3-319-65052-4_2"
}

@article{haakma2014belief,
  title={Belief elicitation to populate health economic models of medical diagnostic devices in development},
  author={Haakma, Wieke and Steuten, Lotte MG and Bojke, Laura and IJzerman, Maarten J},
  journal={Applied health economics and health policy},
  volume={12},
  pages={327--334},
  year={2014},
  publisher={Springer}
}

@article{hemming2012bayesian,
  title={Bayesian cohort and cross-sectional analyses of the PINCER trial: a pharmacist-led intervention to reduce medication errors in primary care},
  author={Hemming, Karla and Chilton, Peter J and Lilford, Richard J and Avery, Anthony and Sheikh, Aziz},
  journal={PloS one},
  volume={7},
  number={6},
  pages={e38306},
  year={2012},
  publisher={Public Library of Science San Francisco, USA}
}

@article{tan2003elicitation,
  title={Elicitation of prior distributions for a phase III randomized controlled trial of adjuvant therapy with surgery for hepatocellular carcinoma},
  author={Tan, Say-Beng and Chung, Y-F Alexander and Tai, Bee-Choo and Cheung, Yin-Bun and Machin, David},
  journal={Controlled clinical trials},
  volume={24},
  number={2},
  pages={110--121},
  year={2003},
  publisher={Elsevier}
}

@article{brown2010effectiveness,
  title={Effectiveness of percutaneous vesico-amniotic shunting in congenital lower urinary tract obstruction: divergence in prior beliefs among specialist groups},
  author={Brown, Celia and Morris, R Katie and Daniels, Jane and Khan, Khalid S and Lilford, Richard J and Kilby, Mark D},
  journal={European Journal of Obstetrics \& Gynecology and Reproductive Biology},
  volume={152},
  number={1},
  pages={25--29},
  year={2010},
  publisher={Elsevier}
}

@article{turner2022borrowing,
  title={Borrowing information across patient subgroups in clinical trials, with application to a paediatric trial},
  author={Turner, Rebecca M and Turkova, Anna and Moore, Cecilia L and Bamford, Alasdair and Archary, Moherndran and Barlow-Mosha, Linda N and Cotton, Mark F and Cressey, Tim R and Kaudha, Elizabeth and Lugemwa, Abbas and others},
  journal={BMC medical research methodology},
  volume={22},
  number={1},
  pages={49},
  year={2022},
  publisher={Springer}
}

@article{diamond2014expert,
  title={Expert beliefs regarding novel lipid-based approaches to pediatric intestinal failure--associated liver disease},
  author={Diamond, Ivan R and Grant, Robert C and Feldman, Brian M and Tomlinson, George A and Pencharz, Paul B and Ling, Simon C and Moore, Aideen M and Wales, Paul W},
  journal={Journal of Parenteral and Enteral Nutrition},
  volume={38},
  number={6},
  pages={702--710},
  year={2014},
  publisher={Wiley Online Library}
}

@article{bojke2021developing,
  title={Developing a reference protocol for structured expert elicitation in health-care decision-making: a mixed-methods study},
  author={Bojke, Laura and Soares, Marta and Claxton, Karl and Colson, Abigail and Fox, Aim{\'e}e and Jackson, Christopher and Jankovic, Dina and Morton, Alec and Sharples, Linda and Taylor, Andrea},
  journal={Health Technology Assessment (Winchester, England)},
  volume={25},
  number={37},
  pages={1},
  year={2021}
}

@article{browne2017bayesian,
  title={A Bayesian analysis of a randomized clinical trial comparing antimetabolite therapies for non-infectious uveitis},
  author={Browne, Erica N and Rathinam, Sivakumar R and Kanakath, Anuradha and Thundikandy, Radhika and Babu, Manohar and Lietman, Thomas M and Acharya, Nisha R},
  journal={Ophthalmic epidemiology},
  volume={24},
  number={1},
  pages={63--70},
  year={2017},
  publisher={Taylor \& Francis}
}

@article{abrams1996bayesian,
  title={A Bayesian approach to Weibull survival models—application to a cancer clinical trial},
  author={Abrams, Keith and Ashby, Deborah and Errington, Doug},
  journal={Lifetime Data Analysis},
  volume={2},
  pages={159--174},
  year={1996},
  publisher={Springer}
}

@article{mason2020framework,
  title={A framework for extending trial design to facilitate missing data sensitivity analyses},
  author={Mason, Alexina J and Grieve, Richard D and Richards-Belle, Alvin and Mouncey, Paul R and Harrison, David A and Carpenter, James R},
  journal={BMC Medical Research Methodology},
  volume={20},
  pages={1--12},
  year={2020},
  publisher={Springer}
}

@article{hiance2009practical,
  title={A practical approach for eliciting expert prior beliefs about cancer survival in phase III randomized trial},
  author={Hiance, Anne and Chevret, Sylvie and L{\'e}vy, Vincent},
  journal={Journal of clinical epidemiology},
  volume={62},
  number={4},
  pages={431--437},
  year={2009},
  publisher={Elsevier}
}

@article{johnson2010valid,
  title={A valid and reliable belief elicitation method for Bayesian priors},
  author={Johnson, Sindhu R and Tomlinson, George A and Hawker, Gillian A and Granton, John T and Grosbein, Haddas A and Feldman, Brian M},
  journal={Journal of clinical epidemiology},
  volume={63},
  number={4},
  pages={370--383},
  year={2010},
  publisher={Elsevier}
}

@article{ren2014assurance,
  title={Assurance calculations for planning clinical trials with time-to-event outcomes},
  author={Ren, Shijie and Oakley, Jeremy E},
  journal={Statistics in Medicine},
  volume={33},
  number={1},
  pages={31--45},
  year={2014},
  publisher={Wiley Online Library}
}

@article{alhussain2020assurance,
  title={Assurance for clinical trial design with normally distributed outcomes: Eliciting uncertainty about variances},
  author={Alhussain, Ziyad A and Oakley, Jeremy E},
  journal={Pharmaceutical Statistics},
  volume={19},
  number={6},
  pages={827--839},
  year={2020},
  publisher={Wiley Online Library}
}

@article{salsbury2024assurance,
  title={Assurance methods for designing a clinical trial with a delayed treatment effect},
  author={Salsbury, James A and Oakley, Jeremy E and Julious, Steven A and Hampson, Lisa V},
  journal={Statistics in Medicine},
  volume={43},
  number={19},
  pages={3595--3612},
  year={2024},
  publisher={Wiley Online Library}
}

@article{mason2017development,
  title={Development of a practical approach to expert elicitation for randomised controlled trials with missing health outcomes: application to the IMPROVE trial},
  author={Mason, Alexina J and Gomes, Manuel and Grieve, Richard and Ulug, Pinar and Powell, Janet T and Carpenter, James},
  journal={Clinical Trials},
  volume={14},
  number={4},
  pages={357--367},
  year={2017},
  publisher={SAGE Publications Sage UK: London, England}
}

@article{white2005eliciting,
  title={Eliciting and using expert opinions about influence of patient characteristics on treatment effects: a Bayesian analysis of the CHARM trials},
  author={White, Ian R and Pocock, Stuart J and Wang, Duolao},
  journal={Statistics in medicine},
  volume={24},
  number={24},
  pages={3805--3821},
  year={2005},
  publisher={Wiley Online Library}
}

@article{holzhauer2022eliciting,
  title={Eliciting judgements about dependent quantities of interest: The SHeffield ELicitation Framework extension and copula methods illustrated using an asthma case study},
  author={Holzhauer, Bj{\"o}rn and Hampson, Lisa V and Gosling, John Paul and Bornkamp, Bj{\"o}rn and Kahn, Joseph and Lange, Markus R and Luo, Wen-Lin and Brindicci, Caterina and Lawrence, David and Ballerstedt, Steffen and others},
  journal={Pharmaceutical Statistics},
  volume={21},
  number={5},
  pages={1005--1021},
  year={2022},
  publisher={Wiley Online Library}
}

@article{sun2013expert,
  title={Expert prior elicitation and Bayesian analysis of the Mycotic Ulcer Treatment Trial I},
  author={Sun, Catherine Q and Prajna, N Venkatesh and Krishnan, Tiruvengada and Mascarenhas, Jeena and Rajaraman, Revathi and Srinivasan, Muthiah and Raghavan, Anita and O'Brien, Kieran S and Ray, Kathryn J and McLeod, Stephen D and others},
  journal={Investigative Ophthalmology \& Visual Science},
  volume={54},
  number={6},
  pages={4167--4173},
  year={2013},
  publisher={The Association for Research in Vision and Ophthalmology}
}

@article{hampson2022improving,
  title={Improving the assessment of the probability of success in late stage drug development},
  author={Hampson, Lisa V and Bornkamp, Bj{\"o}rn and Holzhauer, Bj{\"o}rn and Kahn, Joseph and Lange, Markus R and Luo, Wen-Lin and Cioppa, Giovanni Della and Stott, Kelvin and Ballerstedt, Steffen},
  journal={Pharmaceutical Statistics},
  volume={21},
  number={2},
  pages={439--459},
  year={2022},
  publisher={Wiley Online Library}
}

@article{rietbergen2011incorporation,
  title={Incorporation of historical data in the analysis of randomized therapeutic trials},
  author={Rietbergen, Charlotte and Klugkist, Irene and Janssen, Kristel JM and Moons, Karel GM and Hoijtink, Herbert JA},
  journal={Contemporary clinical trials},
  volume={32},
  number={6},
  pages={848--855},
  year={2011},
  publisher={Elsevier}
}

@article{thirard2020integrating,
  title={Integrating expert opinions with clinical trial data to analyse low-powered subgroup analyses: a Bayesian analysis of the VeRDiCT trial},
  author={Thirard, Russell and Ascione, Raimondo and Blazeby, Jane M and Rogers, Chris A},
  journal={BMC Medical Research Methodology},
  volume={20},
  pages={1--10},
  year={2020},
  publisher={Springer}
}

@article{moatti2013modeling,
  title={Modeling of experts’ divergent prior beliefs for a sequential phase III clinical trial},
  author={Moatti, Marion and Zohar, Sarah and Facon, Thierry and Moreau, Philippe and Mary, Jean-Yves and Chevret, Sylvie},
  journal={Clinical Trials},
  volume={10},
  number={4},
  pages={505--514},
  year={2013},
  publisher={SAGE Publications Sage UK: London, England}
}

@article{see2012prior,
  title={Prior elicitation and Bayesian analysis of the Steroids for Corneal Ulcers Trial},
  author={See, Craig W and Srinivasan, Muthiah and Saravanan, Somu and Oldenburg, Catherine E and Esterberg, Elizabeth J and Ray, Kathryn J and Glaser, Tanya S and Tu, Elmer Y and Zegans, Michael E and McLeod, Stephen D and others},
  journal={Ophthalmic epidemiology},
  volume={19},
  number={6},
  pages={407--413},
  year={2012},
  publisher={Taylor \& Francis}
}

@article{desai2021prior,
  title={Prior elicitation of the efficacy and tolerability of Methotrexate and Mycophenolate Mofetil in Juvenile Localised Scleroderma},
  author={Desai, Yasin and Jaki, Thomas and Beresford, Michael W and Burnett, Thomas and Eleftheriou, Despina and Jacobe, Heidi and Leone, Valentina and Li, Suzanne and Mozgunov, Pavel and Ramanan, Athimalaipet V and others},
  journal={AMRC open research},
  volume={3},
  pages={20},
  year={2021}
}

@article{vail2001prospective,
  title={Prospective application of Bayesian monitoring and analysis in an ‘open’randomized clinical trial},
  author={Vail, Andy and Hornbuckle, Janet and Spiegelhalter, David J and Thornton, Jim G},
  journal={Statistics in medicine},
  volume={20},
  number={24},
  pages={3777--3787},
  year={2001},
  publisher={Wiley Online Library}
}

@article{brownstein2019role,
  title={The role of expert judgment in statistical inference and evidence-based decision-making},
  author={Brownstein, Naomi C and Louis, Thomas A and O’Hagan, Anthony and Pendergast, Jane},
  journal={The American Statistician},
  volume={73},
  number={sup1},
  pages={56--68},
  year={2019},
  publisher={Taylor \& Francis}
}

@article{jansen2024uk,
  title={The UK resuscitative endovascular balloon occlusion of the aorta in trauma patients with life-threatening torso haemorrhage: the (UK-REBOA) multicentre RCT},
  author={Jansen, Jan O and Hudson, Jemma and Kennedy, Charlotte and Cochran, Claire and MacLennan, Graeme and Gillies, Katie and Lendrum, Robbie and Sadek, Samy and Boyers, Dwayne and Ferry, Gillian and others},
  journal={Health Technology Assessment (Winchester, England)},
  volume={28},
  number={54},
  pages={1},
  year={2024}
}

@article{cornelius2024treating,
  title={Treating severe paediatric asthma with mepolizumab or omalizumab: a protocol for the TREAT randomised non-inferiority trial},
  author={Cornelius, Victoria and Babalis, Daphne and Carroll, William D and Cunningham, Steven and Fleming, Louise and Gaillard, Erol and Gupta, Atul and Janani, Leila and Kennington, Erika and Murray, Clare and others},
  journal={BMJ open},
  volume={14},
  number={8},
  pages={e090749},
  year={2024},
  publisher={British Medical Journal Publishing Group}
}

@article{IDEA,
  title={A practical guide to structured expert elicitation using the IDEA protocol},
  author={Hemming, Victoria and Burgman, Mark A and Hanea, Anca M and McBride, Marissa F and Wintle, Bonnie C},
  journal={Methods in Ecology and Evolution},
  volume={9},
  number={1},
  pages={169--180},
  year={2018},
  publisher={Wiley Online Library}
}
\end{document}